\documentclass[twocolumn]{jpsj3}

\usepackage[dvipdfmx]{graphicx}
\usepackage{amsmath}
\usepackage{braket}
\usepackage{bm}
\usepackage{amssymb,amsthm}
\usepackage{float}
\usepackage{color}

\begin{document}

\title{
Angular-time evolution and edge-spin dynamics in the Haldane phase of the $S=1$ bilinear-biquadratic chain
}%

\author{Takuma Kaise and Kouichi Okunishi$^{1,2}$}
\inst{Graduate School of Science and Technology, Niigata University, Igarashi 2, Niigata 950-2181, Japan\\
${}^1$Department of Physics, Osaka Metropolitan University, 3-3-138 Sugimoto, Osaka 558-8585, Japan\\
${}^2$Nambu Yoichiro Institute of Theoretical and Experimental Physics (NITEP), Osaka Metropolitan University, 3-3-138 Sugimoto, Osaka 558-8585, Japan}
\date{\today}%

\abst{
We investigate the angular-time evolution ---a parameter-time evolution generated by the entanglement Hamiltonian--- for the bipartitioned ground state of the S=1 bilinear-biquadratic chain under the open boundary condition with the up edge spin. 
Using a matrix-product-state representation of the ground-state wavefunction, we calculate the angular-time spin correlation functions $\langle S_n^{\alpha }(\tau )S_{n'}^{\alpha }(0)\rangle$ in the Haldane phase, and extract its dominant oscillation mode attributed to the nearly two-fold-degenerate entanglement spectrum associated with the $\mathbb{Z}_2 \times \mathbb{Z}_2$ symmetry.
We also compute the effective edge-spin dynamics under a uniform magnetic field applied to the system part and numerically verify its correspondence to the dominant angular-time mode by precisely comparing the subsystem-size dependence of their amplitudes.}
\maketitle

\section{Introduction}
The ground state of the $S=1$ Heisenberg spin chain\cite{HaldanePLA,Haldane1983,Affleck1989} has long served as a central theoretical platform for understanding a variety of nontrivial concepts in quantum many-body systems.
Starting from the Haldane gap problem for $S=$integer spin chains\cite{Haldane1983,Affleck1989}, several intrinsic features of the $S=1$ Heisenberg chain, such as string-order parameter\cite{denNijs1987}, the effective $S=1/2$ edge spin degrees of freedom, and $\mathbb{Z}_2 \times \mathbb{Z}_2$ symmetry breaking \cite{KT_cmp1992,KT_prb1992},  were established so far.
The explicit construction of the valence-bond-solid (VBS) state, which was initially introduced as the exact ground state for the Affleck-Kennedy-Lieb-Tasaki (AKLT) chain\cite{AKLT,AKLT2}, uncovered a more direct relation between quantum information physics and matrix product states (MPS)\cite{Fannes1989, Fannes1992, Klumper1993};
The auxiliary $S=1/2$ spin degrees of freedom in the VBS state can also be regarded as the successive embedding of Bell pairs, clarifying the quantum information structure inherent in the MPS formalism and providing a foundation for modern understanding of the symmetry-protected-topological (SPT) ordered state\cite{WenRMP2017,Pollmann2010,Liu2011,XieGuWen2011}.
In particular, the degeneracy structure of the entanglement spectrum\cite{Li2008} offers a practical diagnosis of the SPT order\cite{Pollmann2010}, by incorporating tensor network techniques such as density matrix renormalization group (DMRG)\cite{White1992}, which is also based on a generalization of the MPS description of the quantum many-body wavefunction.\cite{Schollwock2011,Orus2019,JPSJ2022,Xiangbook}
Experimentally, the $S=1$ Heisenberg chain was realized as quasi-one-dimensional materials, where the Haldane gap was directly observed\cite{Renard_1987, Katsumata1989}. 
Moreover, the edge spin modes for the open boundary systems were confirmed through electric spin resonance experiments.\cite{DateKindo1990,Hagiwara1990,YoshidaESR}

From the condensed-matter perspective, however, entanglement entropy and entanglement spectrum are usually difficult to access directly, since they are not conventional observables in realistic experimental settings. 
In this paper, we discuss the angular-time-evolution mechanism to capture the entanglement spectrum associated with the Haldane state of the bilinear-biquadratic (BLBQ) chain, which interpolates between the $S=1$ Heisenberg chain and the AKLT chain.
The angular-time evolution is defined as a parameter-time evolution governed by the entanglement Hamiltonian for the bipartitioning of the total system, analogous to the Unruh effect, in which a constantly accelerating observer can detect a thermalized spectrum associated with quantum fluctuations in the vacuum state.\cite{Fulling1973, Unruh1976, RMP_Unruh}
So far, the spin-chain version of the Unruh effect has been proposed for the bipartite partitioning of the XXZ chain based on integrability.\cite{Okunishi2019} 
Then, the angular-time evolution approach has been generalized to the AKLT chain through the entanglement Hamiltonian, enabling an analytic calculation of the autocorrelation function of spins via the MPS formulation.\cite{Nakajima2022}
Furthermore, the gauge transformation for MPS has revealed that the dynamics associated with the effective edge spin under a weak magnetic field applied to the system part mimics the angular-time evolution, capturing the splitting of the degenerate entanglement spectrum due to the open boundary effect in the VBS state.

Our motivation for investigating the BLBQ chain is to extend the edge-spin protocol based on the MPS-gauge transformation for detecting the entanglement spectrum in the VBS state toward the Heisenberg point, where several quasi-one-dimensional materials were actually synthesized.
In contrast to the AKLT point, the entanglement spectrum of the BLBQ chain exhibits a more intricate structure.
The bond dimension of MPS tensors is no longer 2, and thus the resulting angular-time evolution may display complex oscillatory behavior.
Nevertheless, the SPT character of the Haldane state ensures the well-established two-fold degeneracy in the bulk entanglement spectrum, which should be precisely encoded in such a complex behavior of the angular-time evolution.
First, we examine how the SPT nature of the Haldane state is reflected in the angular-time evolution away from the AKLT point.
By using the MPS representation of the ground-state wavefunction computed with DMRG, we numerically extract the most dominant mode  associated with the $\mathbb{Z}_2\times \mathbb{Z}_2$ symmetry from angular-time spin correlation functions.
Next, by comparing this mode with the effective edge spin dynamics under a uniform magnetic field applied to the system part, we numerically verify the correspondence between the edge spin protocol and the angular-time evolution, both of which can describe the splitting of the entanglement spectrum due to the open boundary condition.

This paper is organized as follows.
In the next section, we briefly review the BLBQ chain and the setup of the angular-time evolution, where the reduced density matrix for the ground-state wavefunction is explicitly constructed in the MPS format.
We also explain the relation between the angular-time evolution and the edge-spin protocol in a system-part magnetic field.
In Sec. 3, we present numerical results of the angular-time spin correlation function and perform its Fourier-mode analysis.
In particular, we extract the most dominant mode corresponding to the nearly two-fold-degenerate entanglement spectrum, which is adiabatically connected to that of the AKLT chain.
We also calculate the real-time spin dynamics under a uniform magnetic field applied to the system part to precisely compare the system-part-size dependences of their amplitudes.
Finally, we summarize our results and discuss their relevance to experiments.

\section{Model and formulation}
\label{Sec2}

\subsection{BLBQ chain}
The $S=1$ BLBQ chain has been extensively studied in the context of the Haldane-gap problem, leading to a well-established ground-state phase diagram.\cite{Fath1991,Bursill1995,Schollwock1996,Okunishi1999,Lauchli2006}. 
The Hamiltonian is defined as
\begin{align}
    \mathcal{H}_\beta &=\sum_{i=1}^{N-1} \bm{S}_{i}\cdot\bm{S}_{i+1}+\beta(\bm{S}_{i}\cdot\bm{S}_{i+1})^{2} \, ,
\label{eq_aklt}
\end{align}
where $\bm{S}_i$ is the $S=1$ spin operators at site $i$, and $N$ denotes the length of the chain.
We assume an open boundary condition, with $N$ sufficiently large compared to the correlation length. 
For $-1 <\beta <1$, the BLBQ chain belongs to the Haldane phase and its phase boundaries at $\beta=\pm 1$ are integrable, characterized by gapless excitations.\cite{Takhtajan1982, Babujian1982, Sutherland1975} 
Moreover, $\beta=0$ and $1/3$ correspond to the $S=1$ Heisenberg chain and the AKLT chain, respectively.
Thus, the BLBQ chain often serves as an essential benchmark model for investigating SPT entanglement and edge spin effects associated with the Haldane state, for which the MPS representation of the ground-state wavefunction is especially effective.
In the previous paper\cite{Nakajima2022}, it was analytically demonstrated that, at the AKLT point, the angular-time evolution and the effective edge-spin protocol precisely capture the splitting of the twofold-degenerate entanglement spectrum due to the open-boundary effect. 
In the following, we investigate how the effective edge-spin protocol constructed at the AKLT point can be extended toward the Heisenberg chain, where the entanglement spectrum exhibits a much more complex structure than in the AKLT case.

\subsection{MPS formulation}

\begin{figure}
\begin{center}
\includegraphics[width=7cm]{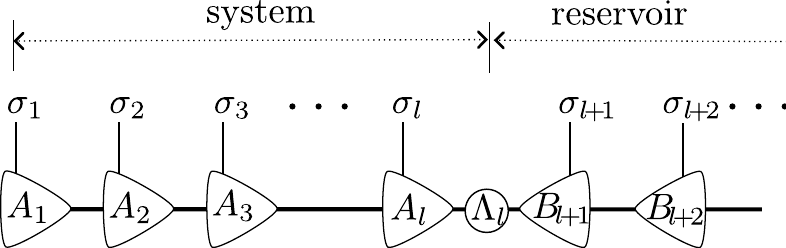}
\end{center}
\caption{
Setup of the system and reservoir parts in the ground-state wavefunction, represented as an MPS in the mixed canonical form.
The system part consists of spins $i=1\cdots l$ from the left edge, while the spins $i>l$ belong to the reservoir part.
A and B respectively denote the MPS tensors obtained by DMRG in the mixed canonical form, and $\Lambda$  represents the singular values at the bond between sites $l$ and $l+1$.
}
\label{tag_fig1}
\end{figure}

In order to investigate the angular-time evolution of the BLBQ chain, we compute its ground state with open boundary conditions using DMRG.
In the Haldane phase, there exist four nearly degenerate wavefunctions corresponding to the singlet and triplet sectors, reflecting the $\mathbb{Z}_{2}\times \mathbb{Z}_{2}$ symmetry.
Here, we focus on the total-$S^z=1$ subspace, which fixes the effective $S=1/2$ edge spins to $\uparrow \uparrow$.
As illustrated in Fig.~\ref{tag_fig1}, we divide the ground-state wavefunction into a system part of length $l(<N)$ and the complemental reservoir part of length $N-l$.
We assume $\xi ,\, l\ll N$, where $\xi$ is the correlation length, implying that the angular-time evolution is essentially governed by the left-edge spin dynamics.

Let us denote the ground state of the BLBQ chain at $\beta$ as $\ket{\Psi_\beta}$.
In conventional DMRG, the ground-state wavefunction is represented as an MPS in the mixed canonical form, where the left singular vectors and right singular vectors meet at the canonical center between $l$ and $l+1$ with the singular-value matrix $\Lambda_l$.
\begin{align}
    \ket{\Psi_\beta} = & \sum_{\{\sigma\}} |\sigma_1 \sigma_2 \cdots \sigma_N \rangle  \nonumber \\
     & \times A^{\sigma_1}_1 A^{\sigma_2}_2 \cdots A^{\sigma_l}_l \Lambda_{l} B^{\sigma_{l+1}}_{l+1} \cdots B^{\sigma_{N}}_{N} \, ,
\label{eq_VBS}
\end{align}
where $\sigma_i \in \{0, \pm 1\}$ denotes the index of an $S=1$ spin state at site $i$.
The diagrammatic representation of Eq.~(\ref{eq_VBS}) is depicted in Fig.~\ref{tag_fig1}, where we have introduced the MPS tensor diagrams:
\begin{align}
 \includegraphics[width=7.5cm]{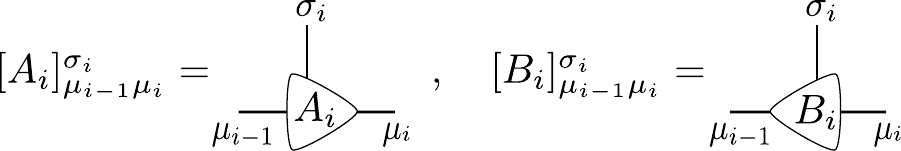}
\end{align}
with the physical spin index $\sigma_i$ and auxiliary indices $\mu_{i-1}$ and $\mu_i$.
Note that in DMRG, the auxiliary index is bounded by the bond dimension $\chi$, and $A$ and $B$ respectively satisfy the left and right canonical conditions, $\sum_\sigma A^{\sigma\dagger} A^\sigma = I$ and $\sum_\sigma B^\sigma B^{\sigma\dagger} = I$, where $I$ is the identity matrix with respect to the auxiliary space.
Also, the boundary tensor elements are explicitly written as $[A_1]^{\sigma_1}_{\mu_1} = \delta_{\sigma_1,\mu_1}$ and $[B_N]^{\sigma_N}_{\mu_{N-1}} = \delta_{\sigma_N,\mu_{N-1}}$.

As in Fig.~\ref{tag_fig1}, we regard the region of length $l$ from the left edge as the system part and its complement as the reservoir part.
By contracting the spins in the reservoir part in Eq.~(\ref{eq_VBS}), we obtain the reduced density matrix for the system part,
\begin{align}
  & \bra{\sigma'_1 \cdots \sigma'_l} \rho_{l} \ket{\sigma_1 \cdots \sigma_l}
   = \!\! \vcenter{\includegraphics[width=4.3cm]{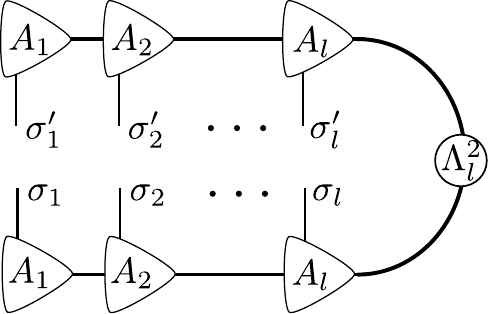}}
  \label{eq_rdm_s1}
\end{align}
Although the matrix size of $\rho_{l}$ is $3^l \times 3^l$, its matrix rank is effectively reduced to $\chi$, which is the bond dimension of the auxiliary degrees of freedom at site $l$.
We can then define the entanglement Hamiltonian from the reduced density matrix
\begin{align}
     \mathcal{K}_{l} \equiv -\log \rho_{l}\, ,
     \label{eq_aHam}
\end{align}
and the corresponding angular-time evolution operator,
\begin{align}
 {\cal U}_{l}(\tau) = e^{-i\tau {\cal K}_{l}}\, ,
\label{eq_atime}
\end{align}
where $\tau$ denotes an angular time generated by $\mathcal{K}_{l}$.

Here, we note that the angular-time evolution was originally introduced as the proper time of a constantly accelerating observer,\cite{Unruh1976,DeWitt,Birrell,Brout,RMP_Unruh} and was later extended to the entanglement Hamiltonian for the bipartition of the XXZ chain as well as the AKLT chain.\cite{Okunishi2019,Nakajima2022}
The entanglement Hamiltonian for quantum spin systems has also been explored in the context of the discretized Bisognano-Wichmann theorem.\cite{Dalmonte2018,Guidici2018}

\subsection{angular-time spin correlations}
On the basis of Eq.~(\ref{eq_atime}), let us consider the angular-time spin correlation function for the BLBQ chain, which is associated with the Unruh-DeWitt detector in a relativistic quantum field theory.
Suppose that $l$ is the length of the system part and $n$ is a site index with $n\le l$.
We define the entanglement Hamiltonian ${\cal K}_{l}$ on the $3^l$-dimensional Hilbert space of the $S=1$ spins in the system part and introduce the angular-time spin correlation function as
\begin{align}
&G^\alpha_{l;n,n'}(\tau) 
      = \mathrm{Tr} [ \rho_{l} S^{\alpha}_{n}(\tau) S^{\alpha}_{n'}(0) ] \, ,
\label{eq_atcorr}
\end{align}
where
\begin{align}
    S^{\alpha}_{n}(\tau) &\equiv e^{i \tau \mathcal{K}_{l}} S^{\alpha}_{n} e^{-i \tau \mathcal{K}_{l}} \, ,
\label{eq_spin1evol}
\end{align}
represents the $S=1$ spin operators with $\alpha \in \{x,y,z\}$ for $1 \le n \le l$ in the Heisenberg representation with respect to the angular time.
Note that Eq.~(\ref{eq_atcorr}) corresponds to the quantum-spin-chain analogue of a scalar-field correlation function in the original Unruh effect in quantum field theory.

For a practical evaluation of Eq.~(\ref{eq_atcorr}), it is convenient to use the spectral decomposition of the entanglement Hamiltonian.
The eigenvalue matrix of $\mathcal{K}_{l}$ is given by
\begin{align}
\Omega_{l} = - 2 \log \Lambda_{{l}} \, .
\end{align}
Here, $\Lambda_{l} = \mathrm{diag}(\lambda_1, \lambda_2, \cdots, \lambda_\chi)$ is the singular value spectrum directly obtained in a DMRG computation.
Then, the entanglement spectrum is explicitly written as
\begin{align}
\Omega_l = \mathrm{diag} ( \varepsilon_1, \varepsilon_2  \cdots, \varepsilon_\chi) 
\end{align}
with $\varepsilon_i \equiv -2 \log \lambda_i$.
The basis change between the physical basis of the system part and the $\chi$-dimensional auxiliary space is given by
\begin{align}
  \langle \sigma_1 \cdots \sigma_l | \mu \rangle
  = A^{\sigma_1}_1 A^{\sigma_2}_2 \cdots A^{\sigma_l}_l \, ,
\label{eq_approx_trans}
\end{align}
where $\ket{\mu}$ denotes the basis that diagonalizes $\rho_{l}$ within the MPS framework.
We emphasize that Eq.~(\ref{eq_approx_trans}) acts as an approximate unitary transformation controlled by the bond dimension $\chi$, in contrast to the AKLT case where Eq.~(\ref{eq_approx_trans}) becomes the exact projection onto the $2 \times 2$ auxiliary space of the VBS state.\cite{Fannes1992,Ostlund1995,Nakajima2022}
Then, the diagrammatic representation of the angular-time spin correlation function is illustrated as
\begin{align}
G^\alpha_{l;n,n'}(\tau) 
    = \label{eq_diagram_atc}
     \!\! \vcenter{\includegraphics[width=4.3cm]{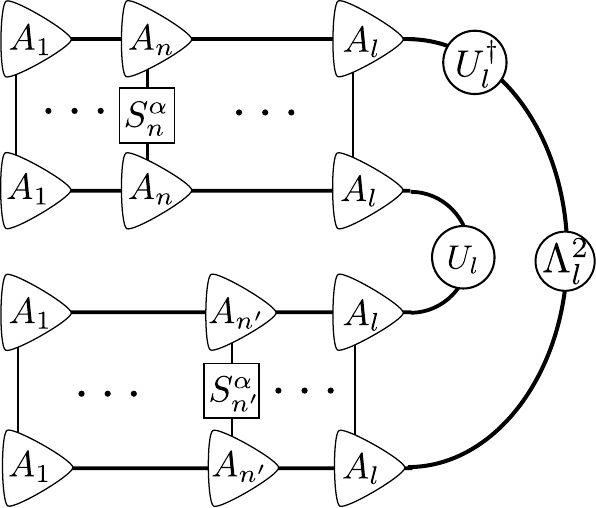}}
    \; ,
\end{align}
where
\begin{align}
U_{l} \equiv e^{-i \tau \Omega_{l}}
\end{align}
denotes the angular-time evolution matrix in the $\ket{\mu}$ basis.
Here, note that, in the diagrammatic representation of Eq.~(\ref{eq_diagram_atc}), we have inserted an approximate identity $I_{\chi }$ between $S_n^\alpha(\tau)$ and $S_{n'}^\alpha(0)$:
\begin{align}
 I_\chi \equiv
    \sum_{\mu=1}^{\chi} 
    \langle \sigma_1 \cdots \sigma_l \mid \mu \rangle 
    \langle \mu \mid \sigma_1 \cdots \sigma_l \rangle \, ,
    \label{eq_incomplete}
\end{align}
which is based on Eq. (\ref{eq_approx_trans}).

In addition to the site-resolved correlation function above, we calculate the angular-time correlation function for the net magnetization in the system part,
\begin{align}
& G^\alpha_l(\tau)
      = \mathrm{Tr} [ \rho_{l} M^{\alpha}_l(\tau) M_l^{\alpha}(0) ] \, ,
\label{eq_m_corr}
\end{align}
where 
\begin{align}
    M^{\alpha}_{l}(\tau) \equiv \sum_{n=1}^{l} S^\alpha_n (\tau) \, .
\label{eq_mevol}
\end{align}
Since $M^{\alpha}_{l}(\tau)$ probes the distribution of the magnetization around the left edge of the chain, 
$G^\alpha_l(\tau)$ would be a more appropriate quantity for experimental observation of the entanglement spectrum.
In practical MPS computations, we evaluate Eq. (\ref{eq_m_corr}) with $G^\alpha_l(\tau) = \sum_{n,n'} G^\alpha_{l;n,n'}(\tau)$.

\subsection{edge spin protocol}

We also analyze the edge-spin protocol for extracting the entanglement spectrum.
At the AKLT point, the angular-time correlation function is equivalent to the autocorrelation function of spins in a weak uniform magnetic field applied to the system part via the ``gauge transformation" for the VBS state.\cite{Nakajima2022}
More precisely, we consider the autocorrelation function of the system-part magnetization, which is defined as
\begin{align}
\mathcal{G}_l^\alpha(\tau) \equiv \langle \mathcal{M}^{\alpha}_{l}(\tau) \mathcal{M}^{\alpha}_{l}(0) \rangle 
\end{align}
where 
\begin{align}
\mathcal{M}^{\alpha}_l(\tau) = e^{-i \tau h_\mathrm{eff}  M^{z}_l  }  M_l^\alpha e^{ i \tau h_\mathrm{eff}  M^{z}_l  } \, ,
\label{eq_effMevo}
\end{align} 
is the magnetization operator in the Heisenberg representation for the system part under an effective magnetic field $h_\mathrm{eff}$.
As noted above, at the AKLT point,  $\mathcal{G}^\alpha_l(\tau)$ mimics the angular-time-evolution dynamics of the effective spin localized near the edge within the subspace of $2\times 2 $ MPS by appropriately choosing $h_\mathrm{eff}= |\varepsilon_{1}-\varepsilon_2|$.
However, such a gauge-transformation-based equivalence is established only at the AKLT point where the ground state is represented by the exact $2\times 2$ MPS.  
Therefore, we numerically examine to what extent the edge spin protocol can be used as a practical tool to describe the angular-time evolution away from $\beta=1/3$.

\section{Numerical analysis}

Using DMRG, we first compute the ground-state wavefunction in the MPS form and then analyze the angular-time correlation function for the system-part magnetization.  
The total chain length is typically $N = 200$, and the subsystem length $l$ is taken up to \(l = 40\), which is sufficiently larger than the ground-state correlation length of the \(S=1\) Heisenberg chain, $\xi \simeq 6.03$ \cite{White1993}.  
Throughout DMRG calculations, we impose open boundary conditions and target the total $S^{z}=1$ sector among the four nearly degenerate ground states.  
Consequently, the effective $S=1/2$ spin at the left edge in Fig.~\ref{tag_fig1} is fixed to $\uparrow$, meaning that time-reversal symmetry is explicitly broken by the boundary condition.  
The bond dimension is taken up to $\chi = 180$, which ensures sufficient convergence of the entanglement spectrum, as shown in Fig.~\ref{tag_fig2}.
In the following, we mainly present results for the $\alpha=x$ component of $G_{l}^\alpha(\tau)$ and $\mathcal{G}_l^\alpha(\tau)$.

\subsection{angular-time correlation function}

\begin{figure}[hbt]
\begin{center}
\includegraphics[width=7.5cm]{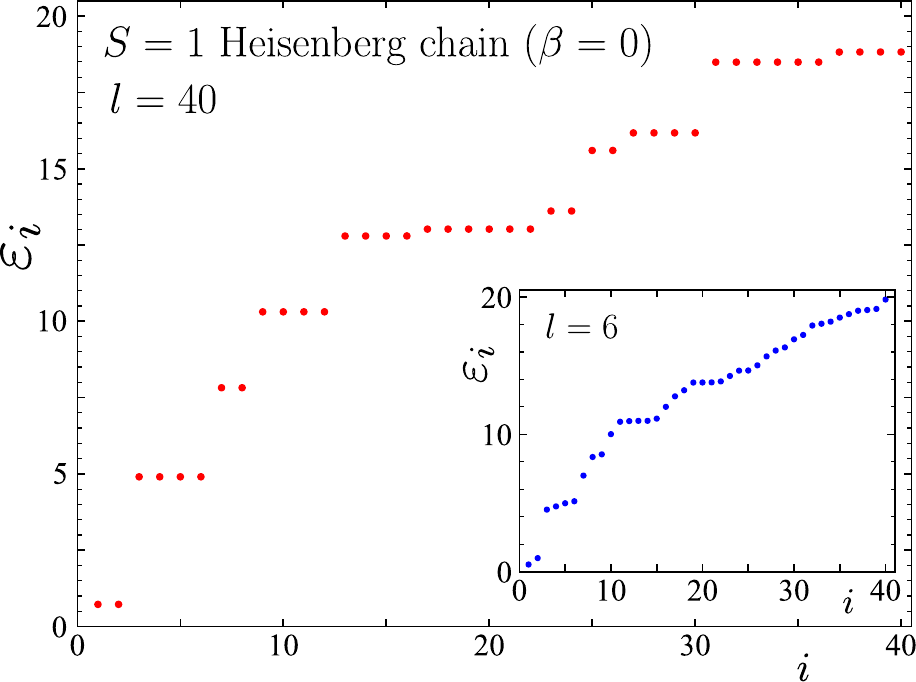}
\end{center}
\caption{
Entanglement spectrum of the bipartitioned ground-state wavefunction of the $S=1$ Heisenberg chain with $N=200$.
The system-part length is $l=40$, which is sufficiently larger than the correlation length.
Accordingly, the spectrum exhibits the characteristic two-fold degeneracy reflecting the bulk behavior.
Inset: Entanglement spectrum for $l=6$, where the splitting of the two-fold degeneracy emerges due to boundary effects
}
\label{tag_fig2}
\end{figure}

\begin{figure}[htb]
\begin{center}
\includegraphics[width=7.5cm]{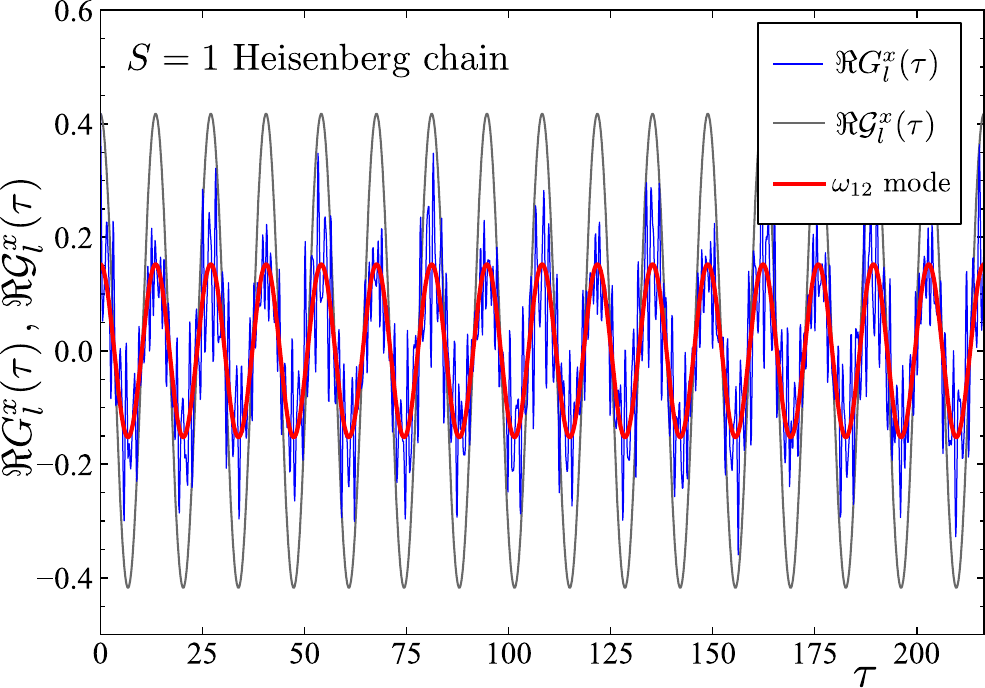}
\end{center}
\caption{
Angular-time correlation function $G_l^x(\tau )$ and the autocorrelation function of the system-part magnetization $\mathcal{G}_{l}^{x}(\tau )$ in the effective magnetic field for the $S=1$ Heisenberg chain with $l=6$ and $N=200$.
The highly oscillatory blue curve and the gray curve represent the real parts of $G_l^x(\tau )$ and $\mathcal{G}_{l}^{x}(\tau )$, respectively.
The red curve shows the $\omega _{12}$ mode extracted from $G_l^x(\tau )$ via Fourier analysis.
}
\label{tag_fig3}
\end{figure}

We evaluate the angular-time correlation function $G_{l}^\alpha(\tau)$ using the diagrammatic expression in Eq. (\ref{eq_diagram_atc}). The quantity exhibits oscillatory behavior with respect to $\tau$ reflecting the entanglement spectrum $\Omega_{l}$.
At the AKLT point ($\beta=1/3$), the matrix rank of the bulk reduced density matrix for the bipartitioning is just two and thus the resulting angular-time correlation function is simply described with a single mode $|\varepsilon_1- \varepsilon_2|=  \log \frac{1+(-1/3)^l}{1-(-1/3)^l}$.
As $\beta $ approaches the Heisenberg point, the effective bond dimension increases, where higher entanglement-spectrum levels descend to the low-energy region.
The entanglement spectra shown in Fig. \ref{tag_fig2} are for the $S=1$ Heisenberg chain.
As seen in the main panel, $\Omega_{l}$ for $l=40$ exhibits two-fold degeneracy originating from the $\mathbb{Z}_2 \times \mathbb{Z}_2$ symmetry for the bulk bipartitioning in DMRG.
However, as shown in the inset of Fig. \ref{tag_fig2}, $\Omega_l$ for $l \lesssim \xi$ displays a clear splitting of the degeneracy due to edge effects. 
These properties of the entanglement spectrum lead to more complex behaviors of the angular-time correlation function when $l$ is relatively small.
A typical example for $\beta=0$ with $l=6(\simeq \xi)$ is presented in Fig. \ref{tag_fig3}, where a superposition of several oscillating modes is observed.
Here, it should be noted that the angular-time spin correlation function may acquire an imaginary component, since the time-reversal symmetry is broken by the fixed edge spin.

In order to extract the intrinsic SPT entanglement involved in the Haldane state, we focus on the lowest entanglement-spectrum levels $\varepsilon_1$ and $\varepsilon_2$, which are adiabatically connected to those of the AKLT chain and thus capture the entanglement associated with the edge spin state.  
For later convenience, we introduce
\begin{align}
\omega_{12} \equiv |\varepsilon_1 - \varepsilon_2|\, .
\end{align}  
Then, an essential point is that the most dominant mode embedded in the complex oscillations of the angular-time correlation function should be described by the $\pm \omega_{12}$ modes.  
We perform a Fourier analysis of the angular-time correlation function, incorporating the Hann window technique, and obtain the Fourier spectra for $l=6$ and $l=40$.  
Details of the Fourier analysis are presented in Appendix A.

\begin{figure}[htb]
\begin{center}
\includegraphics[width=7.5cm]{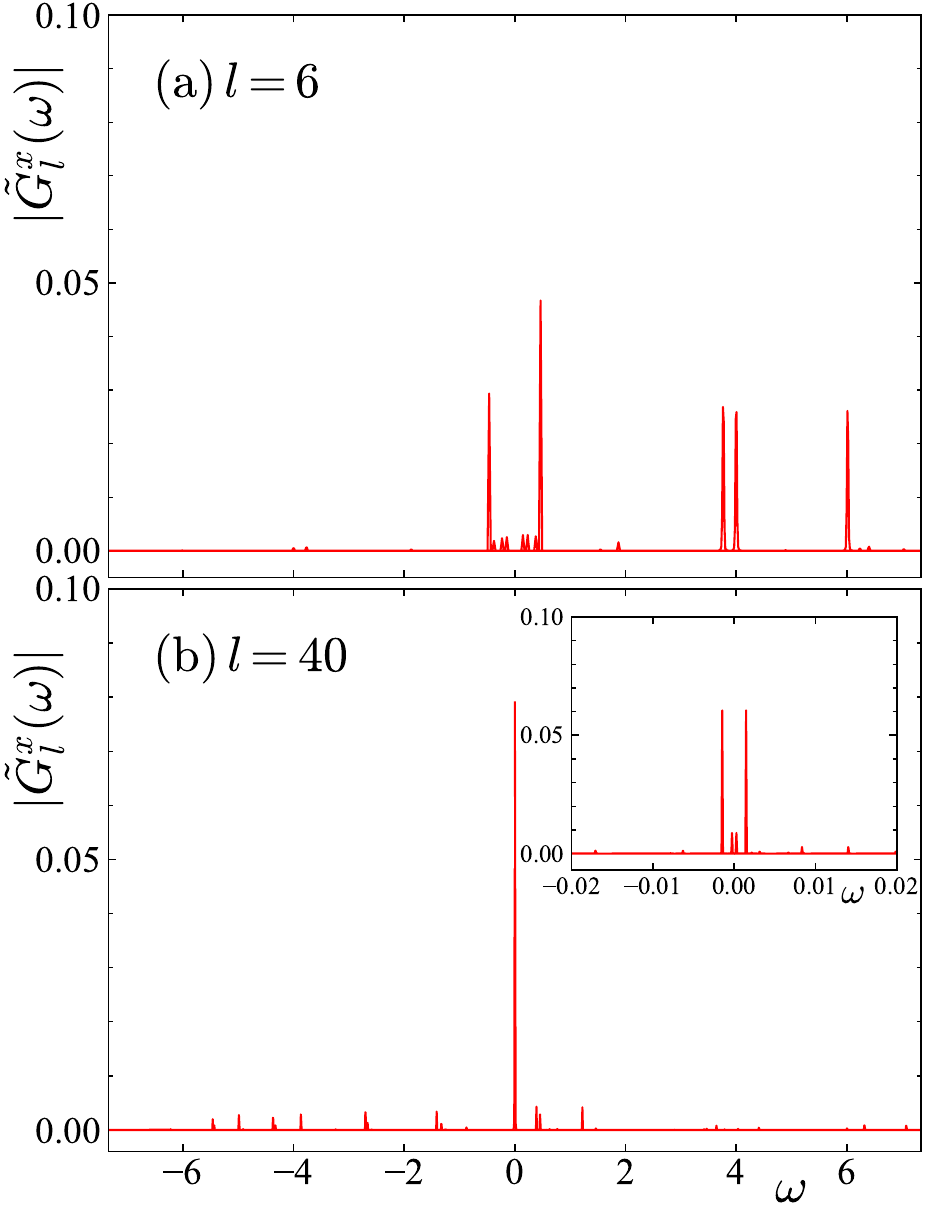}
\end{center}
\caption{
Fourier spectra of $G_l^x(\tau)$ for the Heisenberg chain with $N=200$, which are obtained by the Fourier transformation with the Hann window.
The length of the system part is (a) $l=6$ and (b) $l=40$.
Inset in (b) is the magnification around $\omega = 0$, where the most dominant peaks are observed at $\pm \omega_{12}$.
Note that in the main panel of (b),  these most dominant peaks $\omega\simeq 0$ merge into a single peak due to limited frequency resolution.
}
\label{tag_fig4}
\end{figure}

The resulting spectra for the Heisenberg point are shown in Fig.~\ref{tag_fig4}, where the tallest peaks appear near $\omega = 0$.  
This implies that the $\pm \omega_{12}$ modes play the dominant role in the angular-time evolution dynamics at the Heisenberg point, although its entanglement spectrum has a complex structure containing higher-energy eigenvalues.  
As shown in Fig.~\ref{tag_fig4}(a) and (b), however, quantitative details depend on the system-part length $l$.  
Compared with the bulk spectrum for $l=40$, the spectrum for $l=6$ has relatively taller peaks also in the higher-energy region, implying that splitting of the bulk-degeneracy and the contribution from higher-energy modes become significant for the smaller $l$.
As $l$ increases, the splitting of $\varepsilon_1$ and $\varepsilon_2$ due to the edge effect becomes narrower, and the contributions of higher-energy levels become less dominant in the Fourier spectrum.  
Figure~\ref{tag_fig4}(b) demonstrates that the bulk entanglement in the ground-state wavefunction of the $S=1$ Heisenberg chain can be essentially described by the $\pm \omega_{12}$ modes.

\subsection{edge spin dynamics in a weak magnetic field}

We next present numerical results of the effective edge-spin protocol based on $\mathcal{G}_{l}^{\alpha }$ under a weak magnetic field applied to the system part.
The gray curve of $\mathcal{G}_{l}^{x}(\tau )$ in Fig.~\ref{tag_fig3} shows the autocorrelation function of the system-part magnetization for $l=6$ in the Heisenberg chain.
In contrast to $G_l^x(\tau )$ discussed in the previous subsection, $\mathcal{G}_{l}^{x}(\tau )$ is characterized by a single mode whose period is determined by the effective magnetic field $h_{\mathrm{eff}}=\omega _{12}$ applied to the system part.
We then expect an equivalence between $\mathcal{G}_{l}^{x}(\tau )$ and the $\omega _{12}$ mode embedded in $G_l^x(\tau )$ away from $\beta =1/3$.
However, it should also be noted that the amplitude of $\mathcal{G}_{l}^{\alpha}(\tau )$ is almost twice that of the $\omega _{12}$ mode in $G_l^{\alpha }(\tau )$.

Here, it is important to recall that, in the calculation of ${G}_{l}^{\alpha}(\tau )$,  an approximate identity $I_{\chi }$ was inserted in both sides of the $M(\tau )$ operator.
Although this approximation accurately represents local operators near the canonical center of DMRG, it neglects a part of the correlations contained in $G_l^x(\tau )$, resulting in a smaller amplitude of $G_l^{\alpha }(\tau )$ compared with that of $\mathcal{G}_{l}^{\alpha}(\tau )$ without insertion of $I_{\chi }$.
At the AKLT point, indeed, we have confirmed that inserting $I_{\chi }$  with $\chi=2$ in the calculation of $\mathcal{G}_l^x(\tau )$ reproduces the same amplitude as the $G_l^x(\tau )$.
Then, a crucial point is that $I_{\chi }$ acts as a projection operator onto the ground-state sector in the system part, where the SPT entanglement associated with the $\mathbb{Z}_{2}\times \mathbb{Z}_{2}$ symmetry is exactly preserved, while the discarded components correspond to short-range bulk entanglement contributions. 
Therefore, we examine in detail the $l$-dependence of their amplitudes around the bulk background to verify the equivalence between $\mathcal{G}_{l}^\alpha(\tau )$ and $G_l^{\alpha }(\tau )$ away from the AKLT point.

\subsection{$l$-dependence of the amplitudes}

For quantitatively illustrating the equivalence of $\mathcal{G}_l^\alpha(\tau)$ with the angular-time evolution, we express the $\pm \omega_{12}$ modes from the Fourier spectrum of $G^x_l(\tau)$ in the form,
\begin{align}
 D^x_l \cos(\omega_{12} \tau + \phi) + i D'^x_l \sin(\omega_{12}\tau+\phi')\, ,
\label{tag_omega12}
\end{align}  
where $ D^x_l$ and $D'^x_l $ respectively denote the real and imaginary parts of the $\omega_{12}$ oscillation.
Note that the red curve in Fig.~\ref{tag_fig3} illustrates Eq. (\ref{tag_omega12}) at the Heisenberg point, where eventually $\phi=\phi'=0$ within numerical accuracy.  
For consistency, we also represent $\mathcal{G}_l^{x}(\tau)$ in the same form as Eq. (\ref{tag_omega12}), and write its amplitude as $\mathcal{D}_l^x$ and $\mathcal{D}'^x_l$.

\begin{figure}[hbt]
\begin{center}
\includegraphics[width=8cm]{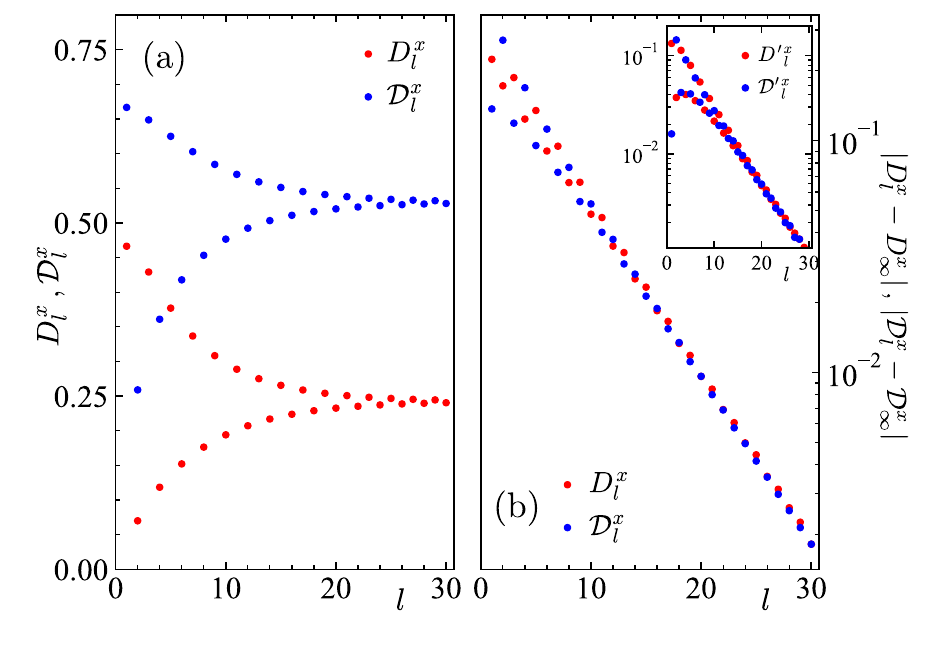}
\end{center}
\caption{
$l$-dependence of the amplitudes:
(a) $D_l^x$ for the $\omega _{12}$ mode (red) and $\mathcal{D}_l^x$ for the autocorrelation function $\mathcal{G}_l^x$ (blue), and
(b) semilog plot of $|D_l^x-D_{\infty }^x|$ (red) and $|\mathcal{D}_l^x - \mathcal{D}_{\infty}^x|$ (blue).
Inset in (b): semilog plot of the amplitudes of the imaginary parts, $D'^x_l$ and $\mathcal{D}'^x_l$.
}\label{tag_fig5}
\end{figure}

We calculate the $l$ dependence of $D_l^x$ and $\mathcal{D}_l^x$ for $0 \le \beta < 1/3$ using the Fourier analysis described in Appendix A.  
The result for the Heisenberg point is shown in Fig.~\ref{tag_fig5}(a). 
Note that the $l$ dependence for other typical $\beta$ are presented in Appendix B. 
As seen in the figure, both $D_l^x$ and $\mathcal{D}_l^{x}$ exhibit oscillatory behaviors around their bulk amplitudes, approximately 0.25 and 0.5, respectively. 
As discussed in the previous subsection, the difference between the two bulk background values arises from the insertion of the projection operator $I_{\chi}$ in the calculation of $G_l^x(\tau)$.  
Meanwhile, the oscillatory behaviors of $D_l^x$ and $\mathcal{D}_l^{x}$ around their bulk amplitudes exhibit remarkably similar patterns, suggesting that both correlation functions are capable of describing the same effective spin dynamics localized near the left edge of the chain.

\begin{figure}[hbt]
\begin{center}
\includegraphics[width=7.5cm]{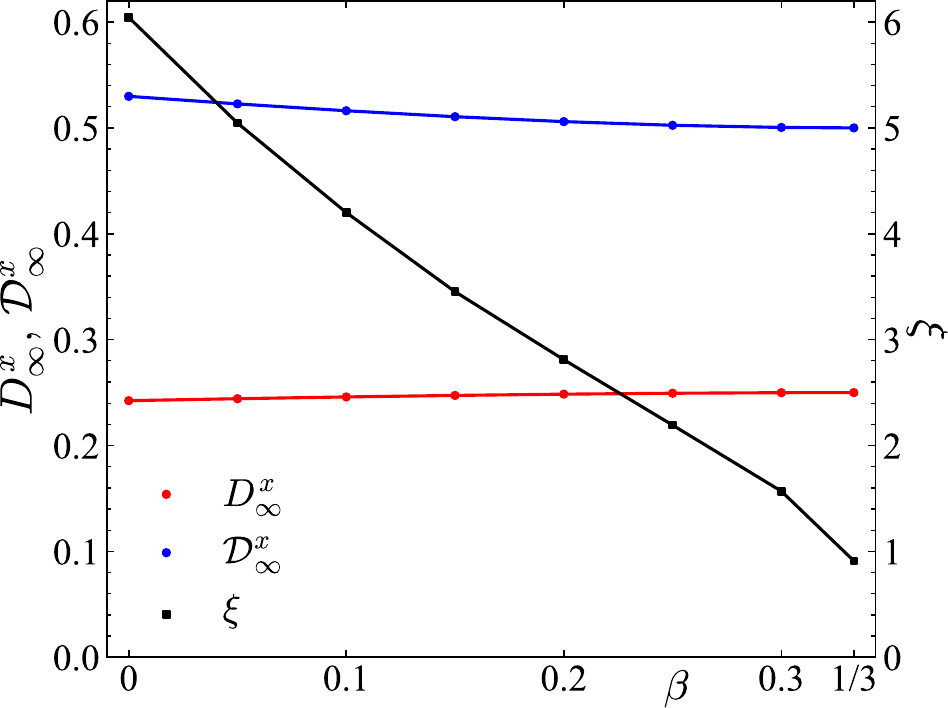}
\end{center}
\caption{
The bulk amplitudes for $D_{\infty}^x$ (red), and $\mathcal{D}^x_{\infty}$ (blue) with $h_\mathrm{eff}=\omega_{12}$.
At $\beta=1/3$, the $\omega_{12}$ mode and  $\mathcal{G}^x_l$ have the exact values 0.25 and 0.5, respectively.
The solid black square indicates the correlation length estimated with Eq. (\ref{eq_D_lfitting}).
}\label{tag_fig6}
\end{figure}

In order to precisely analyze the decay of the amplitudes around their bulk values, we fit the data to  
\begin{align}
D_l^x = a (-)^l e^{-l/\xi} + D_{\infty}^x \, ,
\label{eq_D_lfitting}
\end{align}
where $a$, $D_{\infty}^x$, and $\xi^{-1}$ are treated as fitting parameters.  
We also assume the same asymptotic of $\mathcal{D}_l^x =  a (-)^l e^{-l/\xi} + \mathcal{D}_{\infty}^x $.  
Here, note that $D_{\infty}^x$ and $\mathcal{D}_\infty^x$ represent the bulk amplitudes, and the extracted value of $\xi$ turns out to be consistent with the correlation length obtained from the conventional ground-state spin correlation function.
For the Heisenberg chain, we obtain $D^x_\infty \simeq 0.2424$ for $D_l^x$ and $\mathcal{D}^x_\infty \simeq 0.5298$ for $\mathcal{D}_l^x$, using the data in the range $l = 21 \sim 30$, where the semilog plot exhibits the most stable linear behavior.  
From the same data sets, we estimate $\xi \simeq 6.04$ for both $D_l^x$ and $\mathcal{D}_l^x$, in agreement with the established value of the correlation length for the $S=1$ Heisenberg chain \cite{White1993}.  
Figure~\ref{tag_fig6} summarizes the $\beta$ dependence of the bulk amplitudes $D^x_\infty$ and the correlation length $\xi$, indicating that the analytic result at the AKLT point ($\beta = 1/3$) is essentially preserved in both $D^x_\infty$ and $\mathcal{D}^x_\infty$ for $\beta < 1/3$.

In order to demonstrate the correspondence between $D_l^x$ and $\mathcal{D}_l^x$, we examine the amplitudes after subtracting the bulk contribution.
Figure \ref{tag_fig5}(b) presents a semilog plot of $|D_l^x-D^x_\infty|$ and  $|\mathcal{D}_l^x-\mathcal{D}^x_\infty|$.  
As seen in the figure, the two quantities collapse onto the same exponential-decay line, indicating that $\mathcal{D}_l^{x}$ captures the same physics as the angular-time-evolution dynamics away from the AKLT point.  
This behavior further suggests that the insertion of $I_{\chi}$ primarily affects the background bulk entanglement, while the SPT entanglement associated with the edge-spin mode remains robust in both $G_l^x(\tau)$ and $\mathcal{G}_l^{x}(\tau)$.  
The same correspondence is also observed for the imaginary part of the amplitude, shown in the inset of Fig.~\ref{tag_fig5}(b).  
We therefore conclude that $\mathcal{G}_l^{x}$ faithfully reflects the $\mathbb{Z}_2 \times \mathbb{Z}_2$ entanglement structure to the same extent as the angular-time-evolution dynamics, even away from the AKLT point.
On the other hand, the deviation of the bulk amplitude $\mathcal{D}_\infty^x$ from  ${D}_\infty^x$ is basically attributed to the background short-range entanglements projected out by $I_\chi$.
However, we also note that as $\beta$ approaches the Heisenberg point, the weight of singular values other than $\lambda_1$ and $\lambda_2$ gradually increases in the ground-state wavefunction, making it difficult to distinguish the contribution of such irrelevant modes embedded in $\mathcal{D}_l^x$ when $I_\chi$ is inserted.


Here, we comment on the behavior of the angular-time evolution in the negative-$\beta$ region.  
For the BLBQ chain, the Haldane phase persists down to $\beta=-1$, where the system reaches a quantum critical point described by the $c=3/2$ conformal field theory (CFT). \cite{Takhtajan1982, Babujian1982} 
Our numerical analysis confirms that the equivalence between $G_{l}^{x}$ and $\mathcal{G}_{l}^{x}$ remains valid at least down to $\beta=-0.3$, where the Haldane gap is approximately $0.23$\cite{Schollwock1996}.  
As $\beta\to -1$, the Haldane gap decreases and the correlation length diverges.  
At the same time, higher-energy levels of the entanglement Hamiltonian descend toward the low-energy region, making it increasingly difficult to resolve the $\omega_{12}$ mode within the DMRG calculation.
For such a situation, a perturbed CFT analysis would be more appropriate.\cite{Cho2017}

\section{Summary and Discussions}

In this work, we have investigated the angular-time evolution ---defined as the parameter-time evolution governed by the entanglement Hamiltonian of the bipartitioned ground-state wavefunction--- and the real-time dynamics of the effective edge spin associated with the $\mathbb{Z}_2 \times \mathbb{Z}_2$ symmetry in the Haldane phase of the $S=1$ bilinear-biquadratic chain. 
Using the matrix-product-state representation, we computed the angular-time correlation function $G_l^x(\tau)$ and the autocorrelation function $\mathcal{G}_l^x(\tau)$ of the corresponding effective edge spin under a weak magnetic field applied to the system part. 
The detailed comparisons of these quantities demonstrate that the edge-spin protocol $\mathcal{G}_l^x(\tau)$ faithfully captures the entanglement structure associated with the $\mathbb{Z}_2 \times \mathbb{Z}_2$ symmetry in the Haldane phase, even away from the Affleck-Kennedy-Lieb-Tasaki (AKLT) point. 
This result suggests a possible route toward experimentally probing SPT entanglement through the real-time dynamics of an effective edge spin in a uniform magnetic field for realistic $S=1$ quantum spin chains.

Nevertheless, we should also note that determining the effective field $h_\mathrm{eff}$, or equivalently the entanglement-spectrum splitting $\omega_{12}$, remains a challenging task for a magnetic-resonance experiment based on the edge-spin protocol, unless the value of $h_\mathrm{eff}$ is known a priori. 
This difficulty arises because the edge-spin protocol, being solely constructed from the ground-state wavefunction, does not contain an intrinsic mechanism for identifying $h_\mathrm{eff}$. 
It is, however, noteworthy that electron-spin-resonance experiments have actually observed signals associated with edge-spin modes for Y$_2$BaNi$_{0.96}$Mg$_{0.04}$O$_5$, which can be modeled as an ensemble of short-length $S=1$ Heisenberg chains with weak magnetic anisotropy.\cite{YoshidaESR, Batista1998, Batista1999} 
These observations suggest the importance of exploring the connection between angular-time evolution and edge-spin dynamics for relatively short spin chains, which may provide more direct access to the entanglement spectrum encoded in the ground-state wavefunction.

Recently, a tomography technique in trapped-ion simulators has demonstrated that the entanglement Hamiltonian of a quantum many-body state can be reconstructed as an effective physical Hamiltonian with spatially nonuniform interactions.\cite{Kokail2021}
Several measurement protocols for accessing the spectrum associated with SPT entanglement have also been proposed for ultracold-atom platforms and quantum-circuit architectures,\cite{ES_Measurement, Choo2018} where the ``real-time'' dynamics generated by entanglement Hamiltonians would, in principle, become experimentally observable.
More recently, the valence-bond-solid state of a finite-length AKLT chain has been realized on a quantum computer through its explicit quantum-circuit construction.\cite{SmithPRX2023}
The Unruh effect for quantum field theories in continuous space times is also mapped to quantum spin systems.\cite{Dalmonte2018,Guidici2018,Kinoshita2025} 
These developments further motivate a systematic studies of how the angular-time evolution governed by the entanglement Hamiltonian can be implemented in such experimentally relevant settings.

\begin{acknowledgment}

This work is supported by Grants-in-Aid for Transformative Research Area "The Natural Laws of Extreme Universe---A New Paradigm for Spacetime and Matter from Quantum Information" (Nos. JP21H05182 and  JP21H05191) from JSPS of Japan, and JST CREST (No. JPMJCR24I1).

\end{acknowledgment}

\appendix

\section{Fourier mode analysis}\label{app_A}

We explain details of the Fourier-mode analysis for the angular-time spin correlation function. 
As shown in Fig. \ref{tag_fig3}, the angular-time correlation function exhibits complex oscillations without a clear period. 
Therefore, we compute the angular-time spin correlation functions over a finite interval $[0, \tau_\mathrm{max}]$, assuming $N_s+1$ discrete data points. 
A naive Fourier transformation over this finite time interval contains artificial scattering at $\tau=0$ and $\tau_\mathrm{max}$. 
In order to suppress background noise originating from such unwanted scattering at the boundaries, we employ filtering based on the Hann window function. 
For a series of $G_l^\alpha(\tau_n)$ with $\tau_n \equiv \Delta \tau \times n$, the Fourier spectrum is given by 
\begin{align} 
\tilde{G}^\alpha_l(\omega_k) = \frac{1}{N_s+1}\sum_{n=0}^{N_s} w(\tau_n) {G}^\alpha_{l}(\tau_n) e^{-i \omega_k \tau_n } \, , 
\end{align}
where $\Delta \tau \equiv \tau_\mathrm{max}/N_s $ denotes the time step. 
The Hann window function is given by 
\begin{align} 
w(\tau) = \sin^2\left( \pi\tau/\tau_\mathrm{max} \right) \, . 
\end{align}

Figure \ref{tag_fig4} shows the Fourier spectrum with the Hann window for the angular-time spin correlation function of the $S=1$ Heisenberg chain, where the subsystem length is $l=6$ and $40$. 
In extracting the low-energy peaks near $\omega=0$, we adopt $N_s=3200$ for $\tau_\mathrm{max} \equiv 32 T_\mathrm{p}$, where $T_\mathrm{p} (= 2\pi / \omega_{12})$ is the period of the most dominant mode estimated from the entanglement spectrum in Fig. \ref{tag_fig2}. 
Meanwhile, we adjust the time range to $\tau_\mathrm{max} = 32 T_\mathrm{p} / 100$ when calculating the Fourier spectrum including the high-frequency region up to $\omega \sim 7.0$. 
Here, we note that the resulting spectrum is well converged with respect to the bond dimension, since the cutoff frequency of the entanglement spectrum with bond dimension $\chi=180$ is $\omega_\mathrm{cf} \sim 30$.

\begin{figure}[htb]
\begin{center}
\includegraphics[width=8cm]{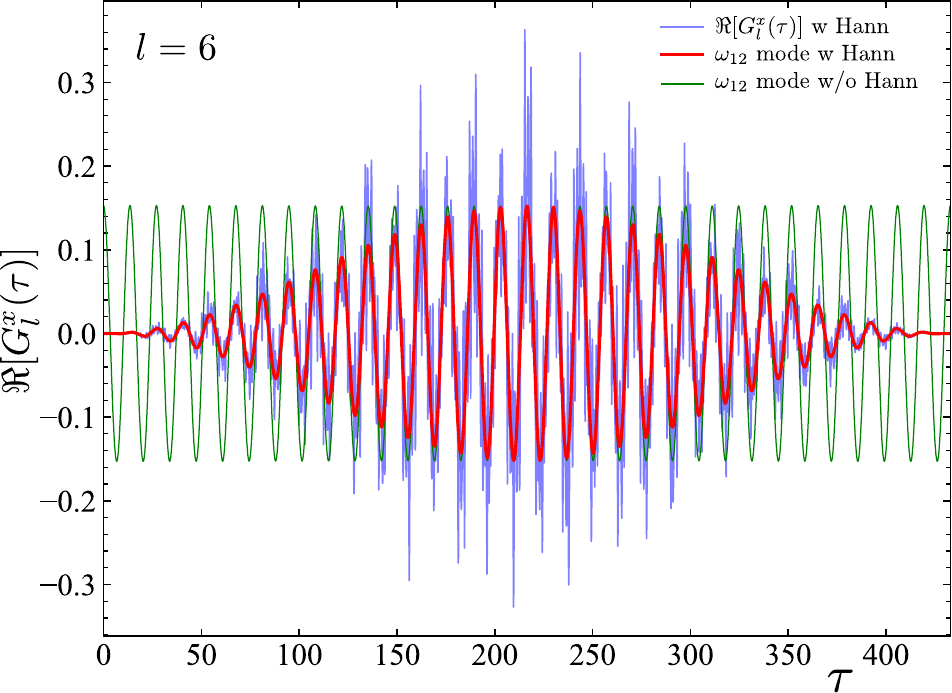}
\end{center}
\caption{
Fourier mode analysis for the angular-time spin correlation function with the Hann window.
The blue curve indicates $\Re[ G_l^x(\tau) ]$ with the Hann window for the $S=1$ Heisenberg chain with $l=6$ and $N=200$.
The red curve indicates  the $\omega_{12}$ mode extracted via a Fourier analysis with the Hann window, which is consistent with the green curve of the $\omega_{12}$ mode  directly estimated without the Hann window.
}
\label{tag_fig_appnd1}
\end{figure}

The peak structure in the Fourier spectra shown in Fig.~\ref{tag_fig4} becomes very sharp when the Hann window is applied.  
Instead, the amplitude of each peak obtained with the Hann window is reduced to almost one half of the original value.  
The correct amplitude can be extracted from the central region of the window function ($\tau \sim \tau_{\mathrm{max}}/2$), where the Hann window does not suppress the signal, as illustrated in Fig.~\ref{tag_fig_appnd1}.  
Although the Fourier spectrum at $\pm \omega_{12}$ are consistent with and without the Hann filtering, the frequency resolution in $\omega$-space is significantly improved when the Hann window is used.  
Figure \ref{tag_fig5} in the main text summarize $l$-dependence of the estimated amplitude at $\beta=0$ represented in the form of Eq. (\ref{tag_omega12}).

\section{$l$-dependence of the amplitudes for various $\beta$}

In the main text, we have mainly focused on numerical results at the Heisenberg point ($\beta=0$). 
Here, we summarize the $l$-dependence of $D_l^x$ for the $\omega_{12}$ mode and $\mathcal{D}_l^x$ for the autocorrelation function of the system magnetization in the range $0 < \beta < 1/3$. 
As $\beta$ approaches $1/3$, the correlation length decreases, allowing us to obtain numerically reliable results with smaller system size $N$ and bond dimension $\chi$ than those required at the Heisenberg point.

\begin{figure}[htb]
\begin{center}
\includegraphics[width=7.5cm]{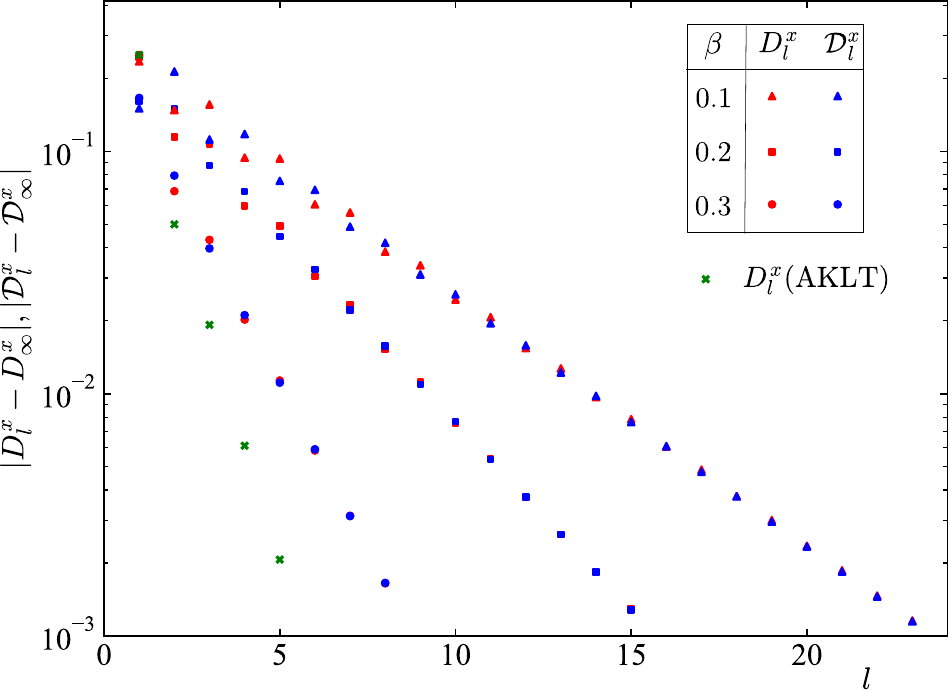}
\end{center}
\caption{
Semilog plots of $|D_l^x - D_\infty^x|$ and $|\mathcal{D}_l^x - \mathcal{D}_\infty^x|$ for $\beta = 0.1$, $0.2$, and $0.3$.
Red and blue symbols represent the numerical results for $D_l^x$ and $\mathcal{D}_l^x$, respectively.
Green crosses indicate the analytic result at the AKLT point.
}
\label{tag_fig_appnd2}
\end{figure}

Figure~\ref{tag_fig_appnd2} shows the resulting $l$-dependence of $|D_l^x - D_\infty^x|$ and $|\mathcal{D}_l^x - \mathcal{D}_\infty^x|$ for $\beta = 0.1$, $0.2$, and $0.3$. 
Note that the bulk values $D_\infty^x$ and $\mathcal{D}_\infty^x$ are presented in Fig.~\ref{tag_fig6}. 
For comparison, we also plot the analytic expression at the AKLT point,
\begin{align}
D_l^x = \frac{1}{4} \frac{1 - (-1/3)^l}{1 + (-1/3)^l},
\end{align}
with $ D_\infty^x = 1/4 $.
The correlation length determined as the inverse decay rate in this figure is also depicted as the solid black square in Fig.~\ref{tag_fig6}, which is in good agreement with the correlation length of the BLBQ chain~\cite{Schollwock1996}. 
These results confirm that the gauge-transformation-based correspondence between the edge-spin protocol and the angular-time evolution in the Haldane phase can be adiabatically connected to the Heisenberg point away from $\beta = 1/3$, even though the entanglement spectrum develops a more complex level structure.

\bibliographystyle{jpsj}
\bibliography{kaise}

\begin{thebibliography}{10}

\bibitem{HaldanePLA}
F.~Haldane: Physics Letters A {\bfseries 93} (1983) 464.

\bibitem{Haldane1983}
F.~D.~M. Haldane: Phys. Rev. Lett. {\bfseries 50} (1983) 1153.

\bibitem{Affleck1989}
I.~Affleck: Journal of Physics: Condensed Matter {\bfseries 1} (1989) 3047.

\bibitem{denNijs1987}
K.~Rommelse and M.~den Nijs: Phys. Rev. Lett. {\bfseries 59} (1987) 2578.

\bibitem{KT_cmp1992}
T.~Kennedy and H.~Tasaki: Communications in Mathematical Physics {\bfseries
  147} (1992) 431 .

\bibitem{KT_prb1992}
T.~Kennedy and H.~Tasaki: Phys. Rev. B {\bfseries 45} (1992) 304.

\bibitem{AKLT}
I.~Affleck, T.~Kennedy, E.~H. Lieb, and H.~Tasaki: Phys. Rev. Lett. {\bfseries
  59} (1987) 799.

\bibitem{AKLT2}
I.~Affleck, T.~Kennedy, E.~H. Lieb, and H.~Tasaki: Comm,Math.Phys. {\bfseries
  115} (1988) 477.

\bibitem{Fannes1989}
M.~Fannes, B.~Nachtergaele, and R.~F. Werner: Europhys. Lett. {\bfseries 10}
  (1989) 633.

\bibitem{Fannes1992}
M.~Fannes, B.~Nachtergaele, and R.~F. Werner: Communications in mathematical
  physics {\bfseries 144} (1992) 443.

\bibitem{Klumper1993}
A.~Kl\"umper, A.~Schadschneider, and J.~Zittartz: Europhys. Lett. {\bfseries
  24} (1993) 293.

\bibitem{WenRMP2017}
X.-G. Wen: Rev. Mod. Phys. {\bfseries 89} (2017) 041004.

\bibitem{Pollmann2010}
F.~Pollmann, A.~M. Turner, E.~Berg, and M.~Oshikawa: Phys. Rev. B {\bfseries
  81} (2010) 064439.

\bibitem{Liu2011}
Z.-X. Liu, M.~Liu, and X.-G. Wen: Phys. Rev. B {\bfseries 84} (2011) 075135.

\bibitem{XieGuWen2011}
X.~Chen, Z.-C. Gu, and X.-G. Wen: Phys. Rev. B {\bfseries 83} (2011) 035107.

\bibitem{Li2008}
H.~Li and F.~D.~M. Haldane: Phys. Rev. Lett. {\bfseries 101} (2008) 010504.

\bibitem{White1992}
S.~R. White: Phys. Rev. Lett. {\bfseries 69} (1992) 2863.

\bibitem{Schollwock2011}
U.~Schollw\"ock: Annals of Physics {\bfseries 326} (2011) 96.

\bibitem{Orus2019}
R.~Or{\'u}s: Nature Reviews Physics {\bfseries 1} (2019) 538–550.

\bibitem{JPSJ2022}
K.~Okunishi, T.~Nishino, and H.~Ueda: Journal of the Physical Society of Japan
  {\bfseries 91} (2022) 062001.

\bibitem{Xiangbook}
T.~Xiang: {\em Density Matrix and Tensor Network Renormalization} (Cambridge
  University Press, 2023).

\bibitem{Renard_1987}
J.~P. Renard, M.~Verdaguer, L.~P. Regnault, W.~A.~C. Erkelens,
  J.~Rossat-Mignod, and W.~G. Stirling: Europhysics Letters {\bfseries 3}
  (1987) 945.

\bibitem{Katsumata1989}
K.~Katsumata, H.~Hori, T.~Takeuchi, M.~Date, A.~Yamagishi, and J.~P. Renard:
  Phys. Rev. Lett. {\bfseries 63} (1989) 86.

\bibitem{DateKindo1990}
M.~Date and K.~Kindo: Phys. Rev. Lett. {\bfseries 65} (1990) 1659.

\bibitem{Hagiwara1990}
M.~Hagiwara, K.~Katsumata, I.~Affleck, B.~I. Halperin, and J.~P. Renard: Phys.
  Rev. Lett. {\bfseries 65} (1990) 3181.

\bibitem{YoshidaESR}
M.~Yoshida, K.~Shiraki, S.~Okubo, H.~Ohta, T.~Ito, H.~Takagi, M.~Kaburagi, and
  Y.~Ajiro: Phys. Rev. Lett. {\bfseries 95} (2005) 117202.

\bibitem{Fulling1973}
S.~A. Fulling: Phys. Rev. D {\bfseries 7} (1973) 2850.

\bibitem{Unruh1976}
W.~G. Unruh: Phys. Rev. D {\bfseries 14} (1976) 870.

\bibitem{RMP_Unruh}
L.~C.~B. Crispino, A.~Higuchi, and G.~E.~A. Matsas: Rev. Mod. Phys. {\bfseries
  80} (2008) 787.

\bibitem{Okunishi2019}
K.~Okunishi and K.~Seki: Journal of the Physical Society of Japan {\bfseries
  88} (2019) 114002.

\bibitem{Nakajima2022}
K.~Nakajima and K.~Okunishi: Phys. Rev. B {\bfseries 106} (2022) 134304.

\bibitem{Fath1991}
G.~F\'ath and J.~S\'olyom: Phys. Rev. B {\bfseries 44} (1991) 11836.

\bibitem{Bursill1995}
R.~J. Bursill, T.~Xiang, and G.~A. Gehring: Journal of Physics A: Mathematical
  and General {\bfseries 28} (1995) 2109.

\bibitem{Schollwock1996}
U.~Schollw\"ock, T.~Jolic\oe{}ur, and T.~Garel: Phys. Rev. B {\bfseries 53}
  (1996) 3304.

\bibitem{Okunishi1999}
K.~Okunishi, Y.~Hieida, and Y.~Akutsu: Phys. Rev. B {\bfseries 59} (1999) 6806.

\bibitem{Lauchli2006}
A.~L\"auchli, G.~Schmid, and S.~Trebst: Phys. Rev. B {\bfseries 74} (2006)
  144426.

\bibitem{Takhtajan1982}
L.~Takhtajan: Physics Letters A {\bfseries 87} (1982) 479.

\bibitem{Babujian1982}
H.~Babujian: Physics Letters A {\bfseries 90} (1982) 479.

\bibitem{Sutherland1975}
B.~Sutherland: Phys. Rev. B {\bfseries 12} (1975) 3795.

\bibitem{DeWitt}
B.~S. DeWitt, Quantum gravity: the new synthesis, In S.~W. Hawking and
  W.~Israel (eds), {\em General Relativity: An Einstein Centenary Survey}.
  Cambridge University Press, Cambridge, 1979.

\bibitem{Birrell}
N.~D. Birell and P.~C.~W. Davies: {\em Quantum Fields in Curved Space}
  (Cambridge University Press, 1982).

\bibitem{Brout}
R.~Brout, S.~Massar, R.~Parentani, and P.~Spindel: Physics Reports {\bfseries
  260} (1995) 329 .

\bibitem{Dalmonte2018}
M.~Dalmonte, B.~Vermersch, and P.~Zoller: Nature Physics {\bfseries 14} (2018)
  827.

\bibitem{Guidici2018}
G.~Giudici, T.~Mendes-Santos, P.~Calabrese, and M.~Dalmonte: Phys. Rev. B
  {\bfseries 98} (2018) 134403.

\bibitem{Ostlund1995}
S.~\"Ostlund and S.~Rommer: Phys. Rev. Lett. {\bfseries 75} (1995) 3537.

\bibitem{White1993}
S.~R. White and D.~A. Huse: Phys. Rev. B {\bfseries 48} (1993) 3844.

\bibitem{Cho2017}
G.~Y. Cho, A.~W.~W. Ludwig, and S.~Ryu: Phys. Rev. B {\bfseries 95} (2017)
  115122.

\bibitem{Batista1998}
C.~D. Batista, K.~Hallberg, and A.~A. Aligia: Phys. Rev. B {\bfseries 58}
  (1998) 9248.

\bibitem{Batista1999}
C.~D. Batista, K.~Hallberg, and A.~A. Aligia: Phys. Rev. B {\bfseries 60}
  (1999) R12553.

\bibitem{Kokail2021}
C.~Kokail, R.~van Bijnen, A.~Elben, B.~Vermersch, and P.~Zoller: Nature Physics
  {\bfseries 17} (2021) 936.

\bibitem{ES_Measurement}
H.~Pichler, G.~Zhu, A.~Seif, P.~Zoller, and M.~Hafezi: Phys. Rev. X {\bfseries
  6} (2016) 041033.

\bibitem{Choo2018}
K.~Choo, C.~W. von Keyserlingk, N.~Regnault, and T.~Neupert: Phys. Rev. Lett.
  {\bfseries 121} (2018) 086808.

\bibitem{SmithPRX2023}
K.~C. Smith, E.~Crane, N.~Wiebe, and S.~Girvin: PRX Quantum {\bfseries 4}
  (2023) 020315.

\bibitem{Kinoshita2025}
S.~Kinoshita, K.~Murata, D.~Yamamoto, and R.~Yoshii: Phys. Rev. Res. {\bfseries
  7} (2025) 043135.

\end{thebibliography}

\end{document}